\documentclass{elsart}
\journal{New Astronomy}

\usepackage{natbib}

\usepackage{graphicx}
\usepackage{epsfig}

\usepackage{amssymb}

\newcommand{\be}{\begin{equation}}
\newcommand{\ee}{\end{equation}}
\def\gtsima{$\; \buildrel > \over \sim \;$}
\def\ltsima{$\; \buildrel < \over \sim \;$}
\def\gsim{\lower.5ex\hbox{\gtsima}}
\def\lsim{\lower.5ex\hbox{\ltsima}}
\def\msun{\,{\rm M_\odot}}
\def\E3{{E}_{\rm g}^{III}}

\begin{document}

\begin{frontmatter}

\title{Detecting Primordial Stars}


\author{Evan Scannapieco$^1$, Andrea Ferrara$^2$, Alexander Heger$^3$}
\author{Piero Madau$^4$ , Raffaella Schneider$^5$,  \&  Stan Woosley$^4$}
\address{$^1$Kavli Institute for
Theoretical Physics, UC Santa Barbara, CA 93106}
\address{$^2$ SISSA/Int. School for Advanced Studies, Via Beirut 4, 34014 Trieste, Italy}
\address{$^3$ Theoretical
Astrophysics Group, T6, MS B227, Los Alamos National Laboratory, Los
Alamos, NM, 87545}
\address{$^4$ Dept. of Astronomy \& Astrophysics, UC Santa
Cruz, Santa Cruz, CA, 95064} 
\address{$^5$ Osservatorio Astrofisico di Arcetri, 
 Largo E. Fermi 5, 50125 Firenze, Italy}

\begin{abstract}

We study the detectability of primordial metal-free stars, both
through  direct searches for their emission, as well as searches for
the resulting supernovae.  We show that  enrichment is a local process
that takes place over an extended redshift range.  While the duration
of the transition from a metal-free to an enriched universe depends on
several unknown factors, in all models late-forming  metal-free  stars
are  found in $10^{7.5}\msun -10^{8.0} \msun$  objects, which are just
large enough to cool within a Hubble time, but small enough to not be
clustered near areas of previous star formation.   We discuss the
observational properties of these objects, some of which may have
already been detected in ongoing surveys of high-redshift
Lyman-$\alpha$ emitters.

If metal-free stars have masses $140 \msun \lsim M_\star \lsim
260\,\msun,$ they are expected to end their lives as
pair-production supernovae (PPSNe), in which an electron-positron
pair-production instability triggers explosive nuclear burning.  Using
the implicit hydrodynamics code KEPLER, we calculate a set of PPSNe
light curves that allows us to assess observational strategies for
finding these objects.  The peak luminosities of typical PPSNe are
only slightly greater than those of Type Ia supernovae, but they
remain bright much longer ($\sim$ 1 year) and have hydrogen lines.
Ongoing supernova searches may soon be able to place stringent limits
on the fraction of very massive stars  out to $z \approx 2.$  The planned
{\em Joint Dark Energy Mission}  satellite  will be able to extend
these limits out to $z \approx 6$.

\end{abstract}


\end{frontmatter}

\section{Introduction}

A long time ago there were no metals. The primordial fireball
generated a whirlwind of neutrinos, an ocean of thermal photons, and a
mountain of light nuclei, but failed to produce elements heavier than
lithium.  In fact, it was not until several hundred million years
later that the universe was first enriched with heavy elements.

Yet, today stellar nucleosynthetic products are everywhere.  The
lowest metallicity galaxies known are enriched to substantial values
of $\sim 0.02 Z_\odot$ \cite{SS72}; the lowest-density regions of the
intergalactic medium (IGM)  appear to be enriched  out to the highest
redshifts probed \cite{S03,A04,Pe03}; and although direct observations
of metal-poor halos stars have uncovered a number unusual abundance
patterns  \cite{Ch02,Ca04}, not a single metal-free star has ever been
detected.

Thus a clear signature of the transition from a metal-free to an
enriched universe has so-far remained unobserved. Furthermore, there
are good reasons to believe that this transition may have marked a
drastic change in the characteristics of stars.  Recent theoretical
studies have shown that collapsing primordial clouds fragmented into
clumps with typical masses $\sim 10^3 \msun$
\cite{Na99,Ab00,Br01,Sc02} and that accretion onto proto-stellar cores
within these clouds was very efficient \cite{Ri02,Ta04}.  The
implication is that the initial mass function (IMF) of  primordial
stars may have been biased to higher masses than observed today.

\section{The Epoch of Metal-Free Stars}

The work discussed in this section and in \S 3 
is described in
further detail in \cite{Sc03}.  As the detailed shape of the
metal-free IMF is highly uncertain, we adopted a wide range of
theoretical models that are based on  studies of primordial gas
cooling and fragmentation \cite{Na99,Ab00,Br01,Sc02}.  In these
investigations, the minimum fragment mass was comparable to  the Jeans
mass at the temperature and density at which molecular hydrogen levels
start to be populated according to LTE, which is $\sim 10^3 \msun$.

With these studies in mind, we parameterized the IMF with  a range of
models in which the IMF is a Gaussian of the form $\frac{1}{\sqrt{2
\pi} \sigma_C} e^{-(M-M_C)^2/2\sigma_C^2},$ where $100 \msun \le M_C
\le 1000 \msun$ and $\sigma^{\rm min}_C \le \sigma_C \le \sigma^{\rm
max}_C$.  Finally, we also explored a ``null hypothesis'' in which the
distribution metal-free stars is given by a Salpeter IMF.  These
models can be parameterized in turn by a single quantity, $\E3$ the
``energy input per unit gas mass,''  which is the product of the
fraction of gas mass that goes into stars ($f_\star$), the number of
supernovae per $\msun$ of stars (${N}_{SN}^{III}$), the kinetic energy
input per SN in units of $10^{51}$ ergs ($E_{\rm SN}^{III}$), and the
fraction of the SN kinetic energy that gets channeled into an outflow,
dispersing the metals ($f_{\rm w}$) outside the host galaxy.  Thus 
$\E3= f_\star {N}_{SN}^{III} E_{\rm SN}^{III} f_{\rm w}$ has units 
of $10^{51}$ ergs $\msun^{-1}.$

Zero-metallicity stars with initial masses $140 \msun \lsim M_\star
\lsim 260\,\msun$ are expected to end their lives as pair-production
supernovae PPSNe, in which an electron-positron pair-production
instability triggers explosive nuclear burning \cite{Bo84,HW02}.  In
this case, the kinetic energy per SN is $\approx 50 \times 10^{51}$
ergs, and the ejected metal mass is $\approx 100 \msun$ \cite{HW02}.
In conventional core-collapse supernovae, the energy per SN is about
$10^{51}$ ergs and roughly 2 $\msun$ of metal are ejected.   Thus
regardless of the IMF, the  total mass in metals is roughly 2 $\msun$
per 10$^{51}$ ergs of kinetic energy input.
 
In other words, despite the many uncertainties surrounding metal-free
stars,  the total energy in each outflow and the total metal mass
dispersed are closely related.   Using a simple 1-D outflow model
\cite{OM88} to study the metallicity evolution within each dispersal
event,  we have found that for the full range of parameters
considered, the mean metallicity of outflows from primordial objects
is well above the critical threshold ($Z_{\rm cr} \sim 10^{-4}
Z_\odot$) that  marks the formation of normal stars \cite{Sc02}.  This
means that primordial star formation continued well past the time at
which the average IGM metallicity reached $Z_{\rm cr}$, as large areas of
the IGM remained pristine while others were enriched to many times
this value.   Enrichment was a local process, and the history of
metal-free formation is likely to have been determined almost
exclusively by  the efficiency of metal ejection,  parameterized by
$\E3$.

\begin{figure}[!t]
\centerline{\psfig{file=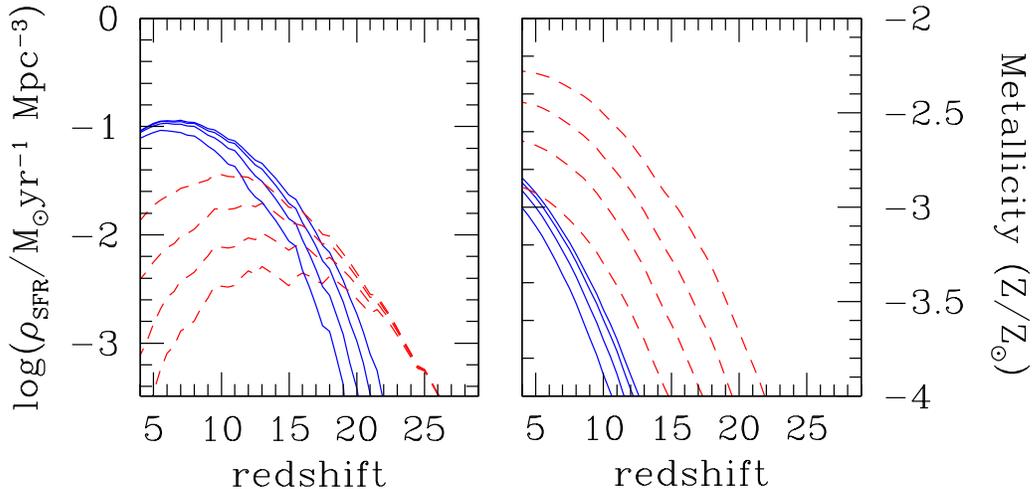,width=5.5in}}
\caption{ Star formation rate densities for primordial and PopII/I
objects.  All models assume $f_\star^{III} = 0.1,$ $f_\star^{II} =
0.1,$ and $f_w = 0.1.$  {\em Left :} SFR density per cubic comoving
Mpc for primordial objects (dashed lines) and PopII/I objects (solid
lines).  We consider models in which, from top to bottom,  $\log_{10}(\E3)$
= $-4.0 $ $-3.5 $, $-3.0 $, and $-2.5 $.  {\em Right:} IGM average
metallicity resulting from primordial (dashed lines) and PopII/I
(solid lines) objects.   Again, $\log_{10}(\E3)$ ranges from $-4.0 $ to
$-2.5$ from  top to bottom.}
\label{fig:sfr}
\end{figure}

To track the dispersal of the first ejected metals,  we made use of
the analytical formalism described in \cite{SB02} [see also
\cite{Po98}] coupled with our 1-D outflow model.  The resulting
evolution of the star formation rate (SFR) density and the mean cosmic
metallicity are shown in Figure~\ref{fig:sfr}.  Here $\E3$
ranges from the $10^{-4}$ value appropriate for a Salpeter IMF  to the
extreme $10^{-2.5}$ value appropriate for an IMF composed almost
solely of $\sim 200 \msun$ stars.

In all models, while the peak of primordial star formation occurs at
$z \sim 10,$  such stars continue to contribute appreciably to the SFR
density at much lower redshifts.  This is true even though the mean
IGM metallicity has moved well past the critical transition
metallicity.  Furthermore, for many models, substantial primordial
star formation continues well into the observable redshift range,
contributing to as much as $10 \%$ of the SFR at $z = 5.$ In all cases
these objects tend to be in the $10^{7.5} \msun - 10^{8.0} \msun$ mass
range, just large enough to cool within a Hubble time, but small
enough that they are not clustered near areas of previous star
formation.

\section{Direct Detections of Primordial Objects}

Employing a simple model of burst-mode star formation, we have
developed a rough characterization of these directly observable
primordial objects.   As metal-free stars are  powerful Ly$\alpha$
line emitters \cite{Tu03,Sch02}, it is natural to use this indicator
as a first step in any search for primordial objects.   The Ly$\alpha$
luminosity,  $L_\alpha,$ is directly related to the  corresponding
hydrogen ionizing photon rate, $Q(H)$, by $L_\alpha = c_L (1-f_{\rm
esc}) Q(H),$ where $c_L \equiv 1.04 \times 10^{-11}$~erg and $f_{\rm
esc}$ is the escape fraction of ionizing photons from the galaxy. For
this uncertain factor we adopt an educated guess of $f_{\rm esc} =
0.2$, which is based on a compilation of  theoretical and
observational results \cite{Ci02}.  We compute $Q(H)$ from
evolutionary stellar models: either from STARBURST99 \cite{Le99} in
the Salpeter IMF case, or from the models in \cite{Sch02} in the case
of very massive stars.

\begin{figure}[!t]
\centerline{\psfig{file=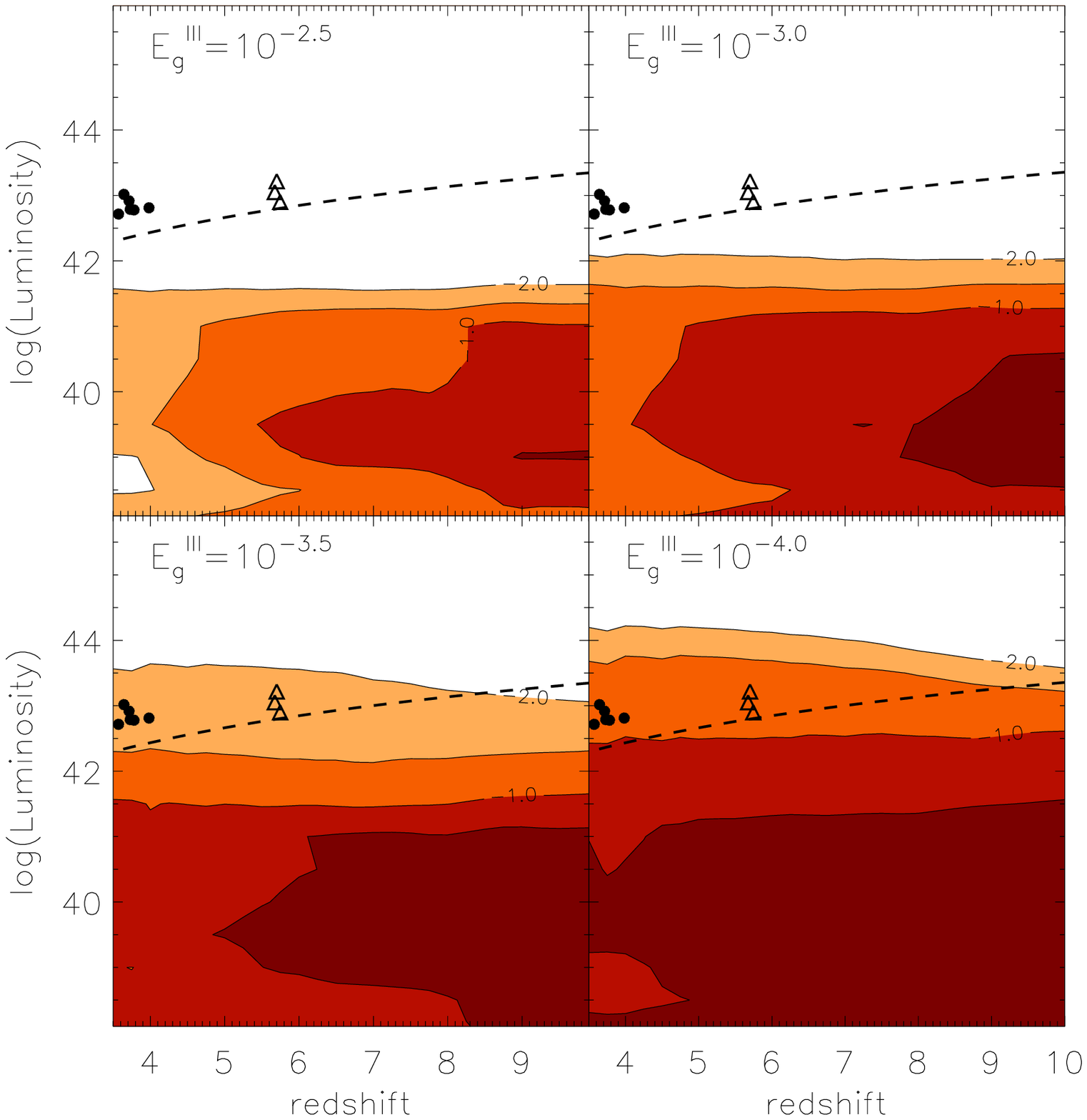,width=2.8in}
\psfig{file=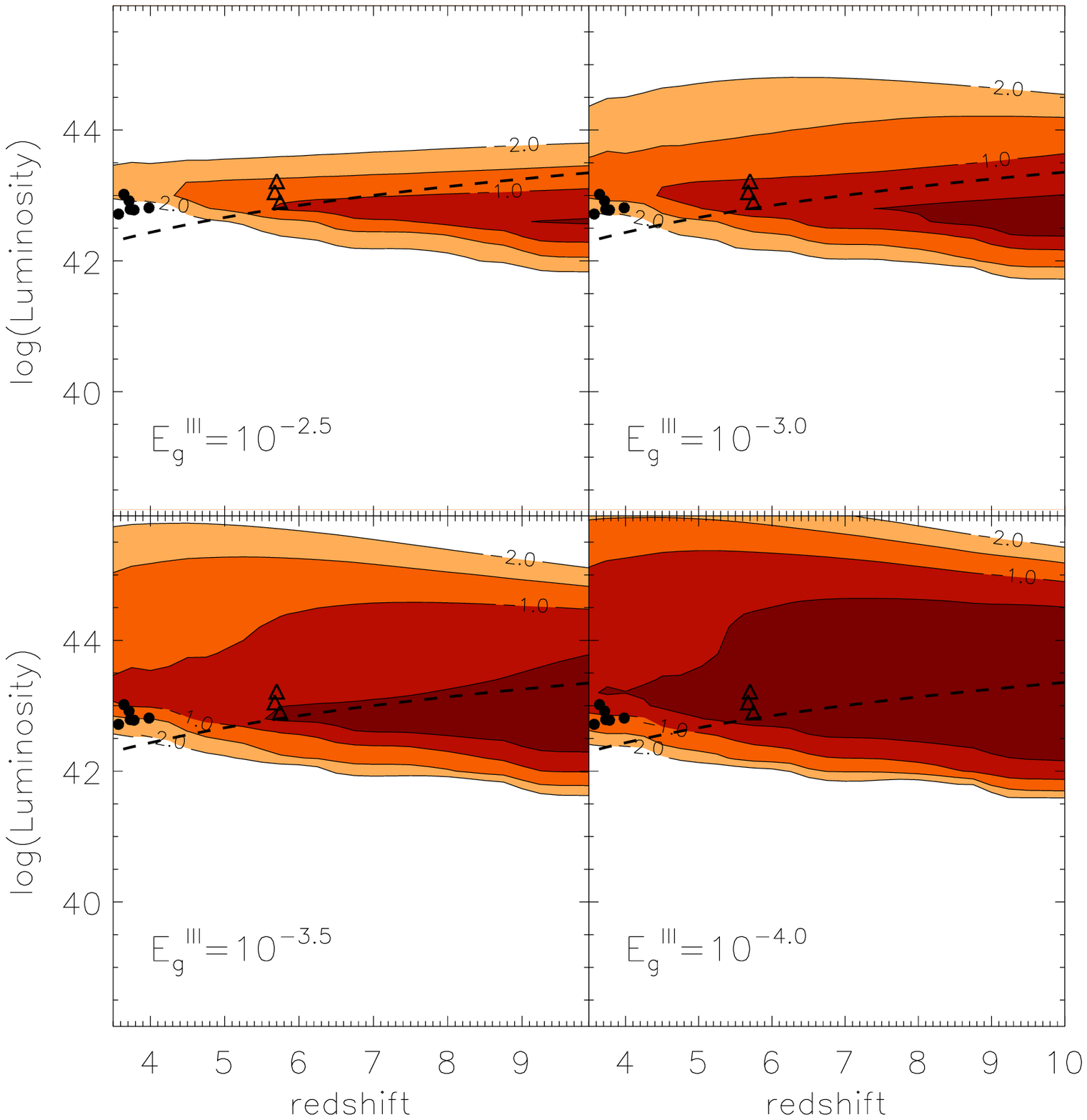,width=2.8in}}
\caption{Fraction of primordial objects as a function of Ly$\alpha$
luminosity and redshift.  Isocontours of fractions $\ge 10^{-2},
10^{-1.5}, 10^{-1},$ and $10^{-0.5}$ are shown.  The left panels
correspond to a model with a Salpeter IMF, while the right panels
correspond to a model in which primordial stars are $\sim 200 \msun$.
Each panel is labeled by its assumed $\E3$ value. For reference,
the dashed line gives the luminosity corresponding to an observed flux
of $1.5 \times 10^{-17}$ ergs cm$^{-2}$ s$^{-1}$, and the points
correspond to a number of recent measurements [See \cite{Sc03} for
details].}
\label{fig:Lya}
\end{figure}

In Figure~2 the isocontours in the Ly$\alpha$ luminosity-redshift
plane indicate the probability to find a galaxy hosting metal-free
stars in a given sample of emitters, again for various values of
$\E3$.  In the left panel of Figure~\ref{fig:Lya}, we consider a model
in which metal-free stars form according to a Salpeter IMF.  Although
in this case the Ly$\alpha$ emission from  these stars is slightly
brighter than from their PopII/I counterparts \cite{Tu03},  this
difference is not dramatic.  However, as outflows are relatively weak
($\E3 \sim 10^{-4}$), and thus $\sim 10\%$ of $z=5$
Ly$\alpha$ emitters with observed fluxes above $1.5 \times 10^{-17}$ ergs 
cm$^{-2}$ s$^{-1}$ are likely be made up of metal-free stars.

In the right panel of Figure~\ref{fig:Lya},  we consider a model in
which the metal-free IMF is peaked at $\sim 200 \msun$.  In this case
the short but bright evolution of such stars boosts their luminosities
by over an order of magnitude.   However, the ejection of the
resulting metals is  dramatically more efficient ($\E3 \sim 10^{-3}$),
and thus again $\sim 10\%$ of $z=5$ Ly$\alpha$ emitters are
metal-free.

Thus it is likely that a number of metal-free objects are lurking at
the limit of present surveys.  The goal then is to find a clear-cut
feature that uniquely identifies a given object as metal-free.  The
most simple such criterion is the equivalent width of Ly$\alpha$,
which should be higher in primordial stars than in a normal
population, as such stars burn hotter and generate more ionizing
photons \cite{Tu03}, particularly in the very-massive case
\cite{Sch02}.    In fact, such large equivalent widths have indeed
been seen in the observed populations of  high-redshifts Ly$\alpha$
emitters \cite{Ma02},  although this can also be the result of other
mechanisms, such as scattering in a clumpy, dusty medium \cite{Ha05}.

A much less ambiguous indicator has been identified by \cite{Tu01},
who noted that primordial stars produce large He III regions, which
may emit detectable He II recombination emission. The recombination
lines at $\lambda$1640 \AA (n = 3 $\rightarrow$ 2), $\lambda$3203 \AA
(n = 5$\rightarrow$ 3), and $\lambda$4686 \AA (n = 4 $\rightarrow$ 3)
are particularly attractive for this purpose because they suffer
minimal effects of scattering by gas and decreasing attenuation by
intervening dust.  For star formation rates of 20 $M_\odot$~yr$^{-1}$
and 5 $M_\odot$~yr$^{-1}$ in a given galaxy, 
the $\lambda$1640 flux is detectable out to
$z \approx 5$ at the sensitivity level of current surveys, although
the flux for $\lambda$4686 is 7.1 times lower.  While these lines also
arise from Wolf-Rayet stars, they would be accompanied by strong metal
lines in this case \cite{AC87}. Metal-free stars, on the other hand, are
uniquely intense sources of pure He-II emission.  Note that this
spectroscopic test is particularly difficult, however, as metal-free
stars should only be found in a small fraction of $z = 5$  objects,
and composite spectra are likely to wipe out such a signal.  Thus the
observed absence of the 1640 doublet in the combined spectra from the {\em
Large Area Lyman Alpha} survey  \cite{Da04} is consistent with all but
the  most optimistic of the models in Fig 2.

\section{Pair-Production Supernovae}

The work discussed in this section 
is described in further detail in \cite{Sc05}.  If primordial stars were
indeed very massive, the resulting pair-production supernovae may be
directly detectable.   This approach has the advantage of constraining
the total rate density of very massive star formation as a function of
redshift, regardless of how they are distributed in galaxies of
different sizes.

\begin{figure}[!t]
\centerline{\psfig{file=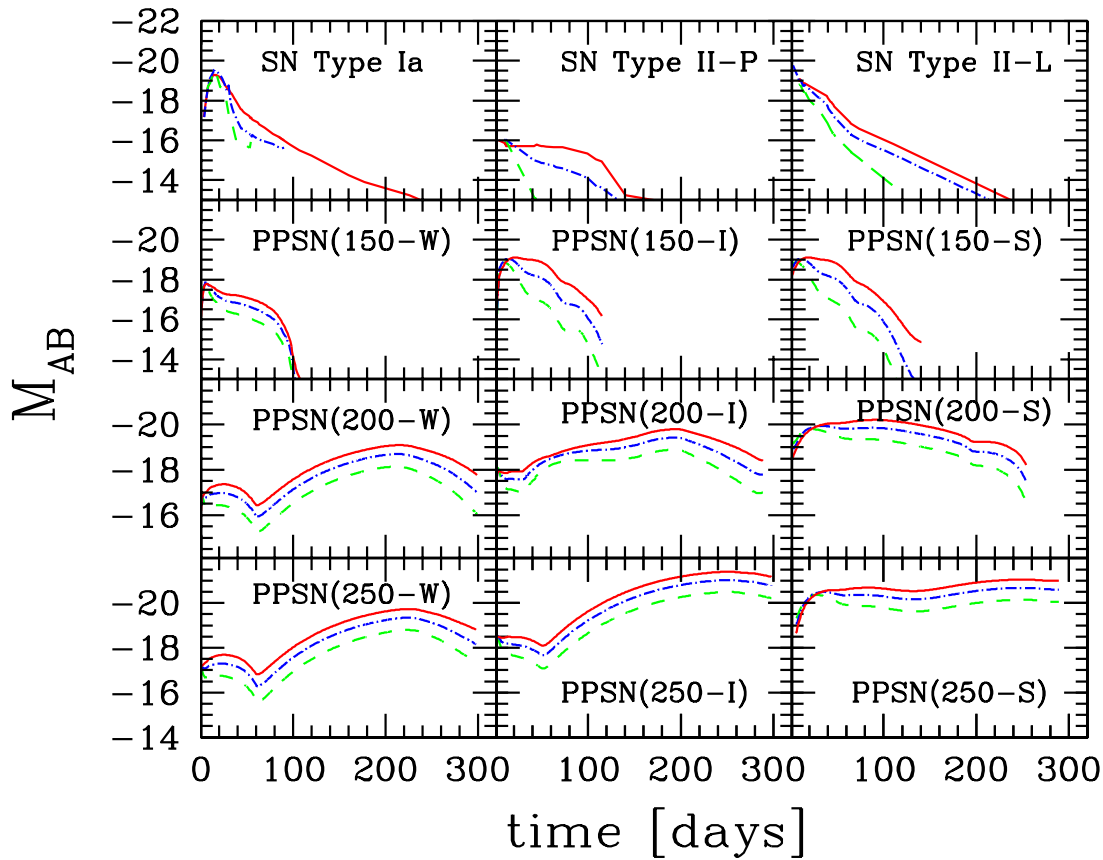,width=6.5in}}
\caption{Comparison of light curves of a SN Type Ia, a SN Type II-P, a
bright SN Type II-L,  and PPSNe models with varying progenitor masses.
In all cases  the solid lines are absolute V-band AB magnitudes, the
dot-dashed lines are the absolute B-band AB magnitudes, and  the
dashed lines are the absolute U-band AB magnitudes. The model names
refer to the mass of the progenitor star (in units of $\msun$) and the
weak (W), intermediate (I), or strong (S) level of  mixing by
convective  overshoot.}
\label{fig:lightcurves}
\end{figure}

Previously, only very limited or approximate models of PPSNe have been
available in the literature \cite{Wo82,He90,He02,Wi05}.  Here we use
the KEPLER code \cite{We78} to directly construct a suite of PPSN
lightcurves that addresses the range of theoretical possibilities and
accounts for the detailed physical processes involved up until the
point at which the SN becomes optically thin.  These results are shown
in  Figure~\ref{fig:lightcurves}.

For most of their lifetimes, the effective temperatures of PPSNe  are
just above the $\approx 10^{3.8} K$ recombination temperature of
hydrogen,  which corresponds to a peak black-body wavelength  $\approx
8000$ \AA.  The most important factors in determining the features in
the luminosity evolution of these SNe are the mass of the progenitor
star and the efficiency of dredge-up  of carbon from the core into the
envelope. In general, increasing the mass leads to greater $^{56}$Ni
production,  which boosts the late time SN luminosity, through
radioactive decay.  Mixing, on the other hand, has two major effects:
it increases the opacity in the envelope, leading to a red giant phase
that increases the early-time luminosity; and it decreases the mass of
the He core, consequently decreasing the  $^{56}$Ni mass and lowering
the late-time luminosity.

Despite these uncertain factors, PPSNe in general can be characterized
as SNe that have all
three key features: (1) peak magnitudes that are brighter than Type
II SNe and comparable or slightly brighter than typical SNe Type Ia;
(2) very long decay times $\approx 1$ year, which result from the  large
initial radii and large masses of material involved in the explosion;
and (3)  the presence of hydrogen lines, which are caused by the outer
envelope.  Note, however, that only this last feature is  present in {\em
all} cases, and in fact,  the lowest mass PPSN models we constructed
have lightcurves that are remarkably similar to those of SN Type II.

\begin{figure}[!t]
\centerline{\psfig{file=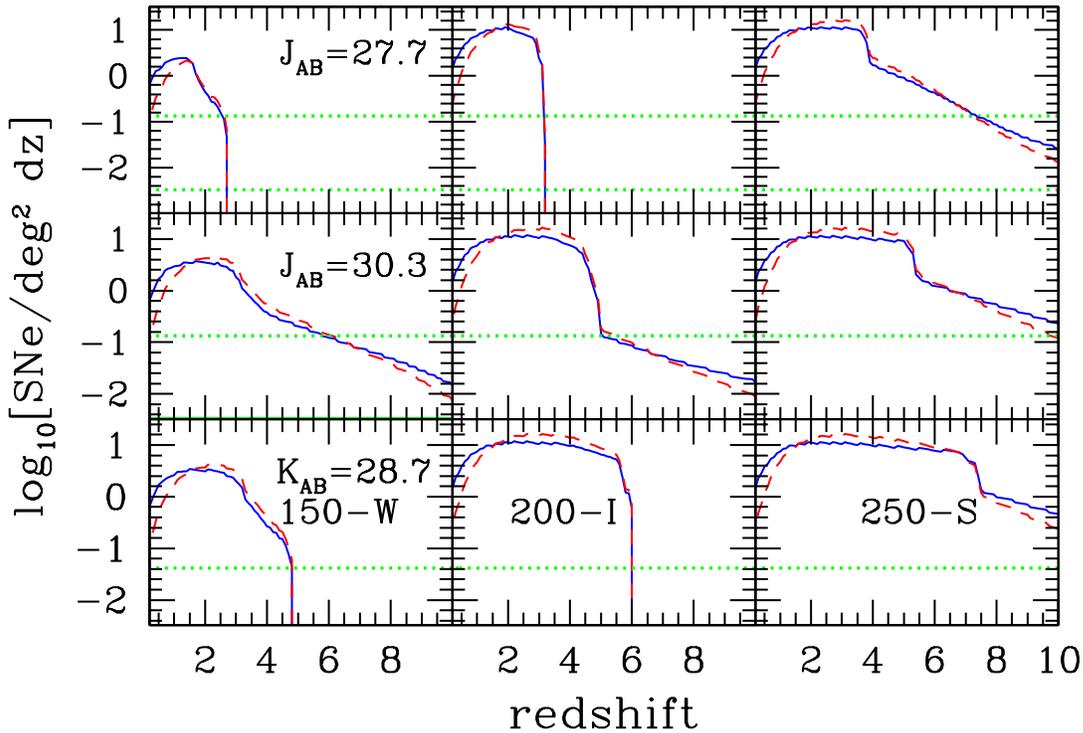,width=6.0in}}
\caption{Number of PPSNe  per square degree per unit redshift above a
fixed broad-band magnitude,  assuming 0.001 $\msun$ yr$^{-1}$
Mpc$^{-3}$ (solid lines) or 1\% of the observed SFR density (dashed
lines). {\em Top:} The $J_{\rm AB} = 27.7$ limit taken in these panels
corresponds to a single scan in the planned deep-field {\em SNAP}
survey, which would cover  7.5 deg$^2$ (indicated by the upper dotted
lines).   A similar magnitude limit with the same coverage would be
obtained with 2 months of grism data from the {\em Destiny} mission.
The full planned 700 deg$^2$ wide-field {\em SNAP} survey  (lower
dotted lines) also has the same limiting magnitude.  {\em Center:} The
$J_{\rm AB} = 30.3$ limit taken in these panels is that of the
deep-field {\em SNAP} survey, which would be able to  place
constraints on PPSN out to $z \gtrsim 5.$ {\em Bottom:} Curves
corresponding to the {\em JEDI} mission, with a magnitude limit of
$K_{\rm AB} = 28.7$ and a coverage of over 24 deg$^2$ with 2 months of
data.  At the highest redshifts, this search  does substantially
better than the fainter, but bluer $J_{\rm AB} = 30.3,$ search.  This
is because at these redshifts the peak of the PPSNe spectra is
shifted to $\approx 40000$ \AA, and  observations at longer
wavelengths represent an exponential increase  in the observed flux.}
\label{fig:counts}
\end{figure}

Accounting for redshifting, time dilation, and appropriate luminosity
factors, we used these lightcurves to relate the overall very massive
SFR density to the number of PPSNe detectable in current and planned
supernova searches.  As the metal-free SFR density is unknown, we
adopted  two simple models, which can be easily shifted  for
comparison  with literature models, including our models in
Fig.~\ref{fig:sfr}.  In the first model, we  assume that metal-free
star formation occurs at a constant rate density, which we take to be
$10^{-3} \, \msun \, {\rm yr}^{-1} \, {\rm Mpc}^{-3},$  independent of
redshift.   In the second case, we assume that at all redshifts
metal-free stars form at 1\% of the observed total SFR density, which
we model  with a simple fit to the most recent measurements
\cite{Gi04,Bo04}.  In all cases we assume that ${N}_{SN}^{III}$, the
number PPSNe per $\msun$ of stars formed, is $10^{-3}.$

Comparing these models with ongoing surveys, we find that significant
limits on PPSNe can already be placed out to moderate redshifts.
Given the area and magnitude limit of the the {\em Institute for
Astronomy  Deep Survey} \cite{Ba04}, for example, one can constrain
very massive star formation to $\lesssim 1 \%$  of the total SFR
density out to a redshift of 2.

Future missions should improve these limits greatly.   In 
Fig.~\ref{fig:counts}, we show the  PPSN constraints that would be obtained
from  three possible realizations of the planned space-based  {\it
Joint Dark Energy Mission (JDEM)}.  
In the upper panel we adopt a limit of $J_{\rm AB} = 27.7$,
corresponding to the wide-field survey planed for the
{\it Supernova Acceleration Cosmology Probe} ({\it SNAP})\footnote{see
http://snap.lbl.gov/} realization  of {\em JDEM}.    With 2 months of
data, the alternative  {\em Dark Energy Space Telescope} ({\em
Destiny})\footnote{see http://destiny.asu.edu/} realization of {\em
JDEM} would cover 7.5 deg$^2$ of sky with spectroscopic observations
down to a similar limit.  In the central panel, we take a limit of
$J_{\rm AB} = 30.3$  which would correspond to the deep-field {\em
SNAP} survey.  In the lower panel, we consider a redder survey with a
limiting magnitude of $K_{\rm AB} = 28.7,$ as appropriate for two
months of observations from  the {\em Joint Efficient Dark-energy
Investigation}   ({\em JEDI})\footnote{see http://jedi.nhn.ou.edu/}
realization of {\em JDEM}.

Although planned with completely different goals in mind, these supernova
searches will also serve as fantastic probes of PPSNe.  Depending on
the realization, {\em JDEM} may be able to place stringent limits on
the rate density of even the faintest PPSNe out to $z \approx 4.$
For more luminous models, PPSNe  will be constrained out to $z \approx
6$ and beyond.  Thus, while a definitive measurement of metal-free
stars  remains allusive, narrow-band, spectroscopic, and SNe searches
are fast  closing in  on these objects. The future of primordial star
detection looks bright.


I would like to thank 
UC Irvine and the organizing committee for
hosting this enjoyable and informative workshop.

\end{document}